\documentclass[letterpaper,prb,preprint, floatfix, showpacs]{revtex4}

\usepackage{epsfig}
\usepackage{subfigure}
\usepackage{amsmath}
\usepackage{amssymb}
\usepackage{chemformula}

\usepackage{bm}
\usepackage{color,soul}
\begin{document}
\title{Tuning the Magnetic and Electronic Properties of Monolayer \ch{VI3} by 3d Transition Metal Doping: A First-Principles Study}
\author{Charles Sun and Xuan Luo}
\affiliation{National Graphene Research and Development Center, Springfield, Virginia 22151, USA}
\date{\today}
\begin{abstract}                                                                                                                                                
\setlength{\parindent}{1cm}

Two-dimensional (2D) materials with robust magnetism have drawn immense attention for their promising applications in spintronics. Recently, intrinsic ferromagnetic vanadium triiodide (\ch{VI_3}) has been synthesized experimentally. To enhance its spintronic property, we modified \ch{VI_3} by interstitial doping with 3d transition metals (TM) and used first-principles calculations to investigate the geometric structure, formation energy, electronic property, and magnetism of pristine  \ch{VI_3} and 3d TM-doped \ch{VI_3} monolayer. Among eight transition metal (Sc-, Ti-, V-, Cr-, Mn-, Fe-, Co-, and Ni-) doped \ch{VI_3} materials, four of them (Ti-, V-, Mn-, and Ni-doped \ch{VI_3}) show robust magnetism with full spin polarization near the Fermi energy. Our research demonstrates that Ti-doped \ch{VI_3} results in half-metallic semiconductor properties (HMS), while V-doped \ch{VI_3} and Ni-doped \ch{VI_3} result in half-semiconductor properties (HSC). Surprisingly, Mn-doped \ch{VI_3} exhibits an unusual bipolar magnetic semiconductor property (BMS). This unique combination of strong ferromagnetism and 100\% spin polarization with a half-metallic, half-semiconductor, or bipolar semiconductor property renders 3d TM-doped \ch{VI_3} as potential candidates for next generation semiconductor spintronic applications. These spin-polarized materials will be extremely useful for spin-current generation and other spintronic applications.

\end{abstract}

\newpage
\maketitle
\section{Introduction}

\begin{figure}[ht]
    \centering
    \includegraphics[width=15cm]{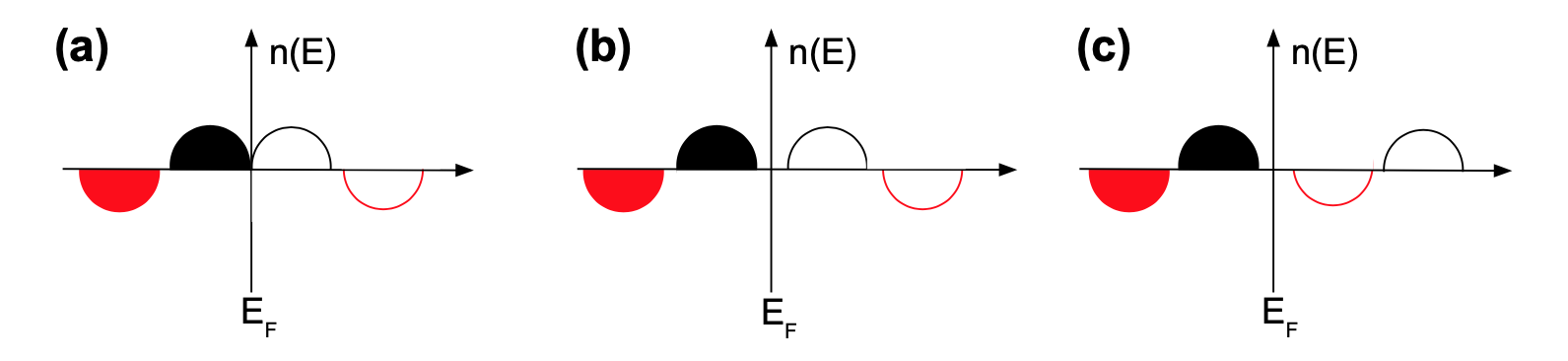}
    \caption{The schemes of density of states n(E) as a function of energy for (a) half-metal semiconductors (HMS), (b) half-semiconductors (HSC), and (c) bipolar magnetic semiconductors (BMS). Black and red represent the spin-up and spin-down states, respectively.
}
    \label{fig:figure1}
\end{figure}

Today’s rapid growth of advanced information technology benefits from smaller and faster semiconductor devices such as transistors. 
Although Moore's Law has successfully predicted since 1965 that the number of transistors would double in integrated circuits every two years, this law no longer holds true due to the physical constraints on the miniaturization of transistors. The size reduction of transistors has increased leakage current and waste heat, both of which create power consumption issues\cite{theis2017end, huang2015moore}. Hence, searching for alternative solutions to overcome this limitation is crucial. Spintronics is an emerging field for next generation nanoelectronic devices that not only reduces power consumption but also increases memory and processing capabilities\cite{hirohata2020review}. Instead of using electron charges, spintronics stores information based on an electron’s spin quantum value: spin-up or spin-down. For next generation semiconductor applications, effective spintronic materials must have an ordered spin structure and full spin polarization. Half-semiconductors (HSC), bipolar magnetic semiconductors (BMS), and half-metallic semiconductors (HMS), also known as spin gapless semiconductors (SGS) \cite{wu2017half, li2016first, wang2008proposal}, meet the criteria because they can intrinsically provide single spin channel electrons with spin polarization reaching 100\% (i.e. full spin polarization) \cite{li2016first}. As shown in Figure \ref{fig:figure1}, HMSs have one conducting spin channel and one semiconducting spin channel. HSCs show semiconducting behavior with a narrow band gap in one spin channel but a wide band gap in the other spin channel. On the other hand, BMSs are characterized by a unique electronic structure in which the valence maximum band (VBM) and conducting minimum band (CBM) are fully spin polarized in the opposite spin direction. To aim for the continuation of semiconductor technology development, seeking for new nanoscale spintronic materials with robust magnetic and diverse electronic properties has become necessary.

Two-dimensional (2D) materials such as graphene\cite{novoselov2005two} hold great potential for their applications of spintronic devices\cite{neto2009electronic}. However, graphene is gapless and possesses no magnetism, which greatly limits its application in the field of spintronics. Although modifications in graphene by a presence of vacancies \cite{ugeda2010missing}, adatoms \cite{krasheninnikov2009embedding,yazyev2007defect}, and nanoribbons\cite{golor2014quantum, wassmann2008structure, zhang2007magnetic} can introduce magnetic properties, intrinsic magnetism did not exist for 2D materials until the report of 
\ch{Cr_2Ge_2Te_6} and \ch{CrI_3} monolayer\cite{gong2017discovery, huang2017layer}. Since then, \ch{CrI_3} has been studied vastly to improve its magnetic property including carrier doping, heterostructures, and Li-adsorption \cite{mcguire2015coupling, zhang2015robust, wang2016doping,huang2017layer, lado2017origin, klein2018probing}. Furthermore, the exfoliation of a transition-metal (TM) halide monolayer is easily achievable due to the weak van der Waals interaction between layers \cite{zhou2016evidencing, he2016unusual, kong2019vi3}. This important finding has also sparked many publications on TM halides \cite{bruyer2016possibility, kulish2017single, mcguire2015coupling, huang2018prediction, huang2017quantum, zhou2015new, ersan2019exploring, valenta2021pressure, carva2021magnetism, bergeron2021polymorphism}.

While many studies have been focused on \ch{CrI_3}, a recent rediscovery \cite{kong2019vi3} of \ch{VI_3} has attracted enormous interest to \ch{VI_3} research. Similar to the \ch{CrI_3} structure, V centers in an octahedral environment with a V honeycomb arrangement in the plane. However, the electron configurations of \ch{Cr^3+} and \ch{V^3+} are different. \ch{Cr^3+} has three valence electrons filled into all three \ch{t_{2g}} electronic states, whereas \ch{V^3+} only partially fills two electrons in the degenerate \ch{t_{2g}} electronic states, which in turn offers more possibilities for tuning. Current simulation studies have indicated that the enhancement of magnetic properties of \ch{VI_3} by strain, interstitial V doping, or substitutional doping of TMs is effective \cite{yang2020enhancement, baskurt2020vanadium, an2019tuning}. In addition, Kong et al. \cite{kong2019vi3} has experimentally revealed vacant interstitial sites within the honeycomb layers that appear to be partially occupied by a small amount of V. Therefore, it is chemically possible to interstitially dope different TM atoms to tune the magnetic and electronic property of \ch{VI_3}. Motivated by these findings, we decided to interstitially dope \ch{VI_3} with eight 3d transition metals so that a systematic study of spintronic property on these materials can be conducted.

In this paper, we use first-principles calculations to investigate the electronic and magnetic properties of 3d TM-doped \ch{VI_3}. We show that \ch{VI_3} not only has excellent electronic and magnetic properties but also demonstrates the immense potential for designing various spintronic materials by tuning its properties to half-metallic semiconductor, half-semiconductor, or bipolar magnetic semiconductor applications. We first describe our methodology for calculating the band gaps and magnetization of these materials. Next, we comprehensively analyze and discuss these results of our calculations for each of the dopant materials. Finally, we summarize the effectiveness of 3d TM doping for semiconductor spintronic applications.

\section{Method}

The materials we studied were pristine \ch{VI_3} and 3d TM (Sc, Ti, V, Cr, Mn, Fe, Co, and Ni)-doped \ch{VI_3}. 

\subsection{Computational Details}

Electronic and magnetic properties were calculated based on the Density Functional Theory (DFT) \cite{kohn1965self, hohenberg1964inhomogeneous}. The simulations were conducted with ABINIT code \cite{gonze2002first} within the Generalized Gradient approximation (GGA), using  the Perdew–Burke–Ernzerhof (PBE) exchange–correlation functional \cite{perdew1996generalized}. The pseudopotentials were described using the projector augmented wave (PAW) method \cite{kresse1999ultrasoft}. To calculate the magnetic properties of materials, spin polarized calculations were utilized. 

\subsection{Convergence and Relaxation}

We converged the plane-wave kinetic energy cutoff, k-point mesh, and vacuum space when the energy difference between consecutive datasets was less than 0.0001 Hartree (about 3 meV) twice. The self-consistent field (SCF) criterion for energy and atom force convergence were set to 1.0 $\times$ 10$^{-10}$ Ha and 1.0 $\times$ $10^{-6}$ Ha/Bohr, respectively. For pristine \ch{VI_3}, the converged plane-wave kinetic energy cutoff of 15 Ha (400 eV), 8 $\times$ 8 $\times$ 1 k-point mesh, and vacuum distance of 11 {\AA} were used. The relaxation terminated when the maximum absolute force on each atom was less than 5.0 $\times$ $10^{-5}$ Ha/Bohr. Then, using spin polarized calculations and the relaxed values, we calculated  the  total  magnetization  and  total  energy  of  the  atomic system and performed band structure and density of states calculations. For bulk TM calculations, the optimized lattice parameters and total energy were calculated using the method above.

For 3d TM-doped \ch{VI_3}, we took the bigger plane-wave kinetic energy cutoff between the two pristine materials (\ch{VI_3} and bulk TM). With the bigger kinetic energy cutoff, k-point mesh, and vacuum distance, we performed a structural relaxation to determine the optimized atomic positions and lattice constants of the material. Then, using spin polarized calculations and these relaxed values, we calculated  the  total  magnetization  and  total  energy  of  the  atomic system and performed band structure and density of states calculations.
\subsection{Electronic Structure}

Band structures were calculated in the first Brillouin zone in the reciprocal space along the high-symmetry k-points of $\Gamma$ (0, 0, 0), M (0.5, 0, 0), K (2/3, 1/3, 0), and $\Gamma$ (0, 0, 0). We also plotted the partial density of states, which show the orbital decomposition of density of states for each atom. 

\section{Results and discussion}

We calculate the electronic and magnetic property of pristine \ch{VI_3} and determine the ground state between its FM and AFM states. We then focus on the results and discussion of 3d TM-doped \ch{VI_3} calculations.

\subsection{Pristine \ch{VI_3} Calculations}

\begin{figure}[htp]
    \centering
    \includegraphics[width=16cm]{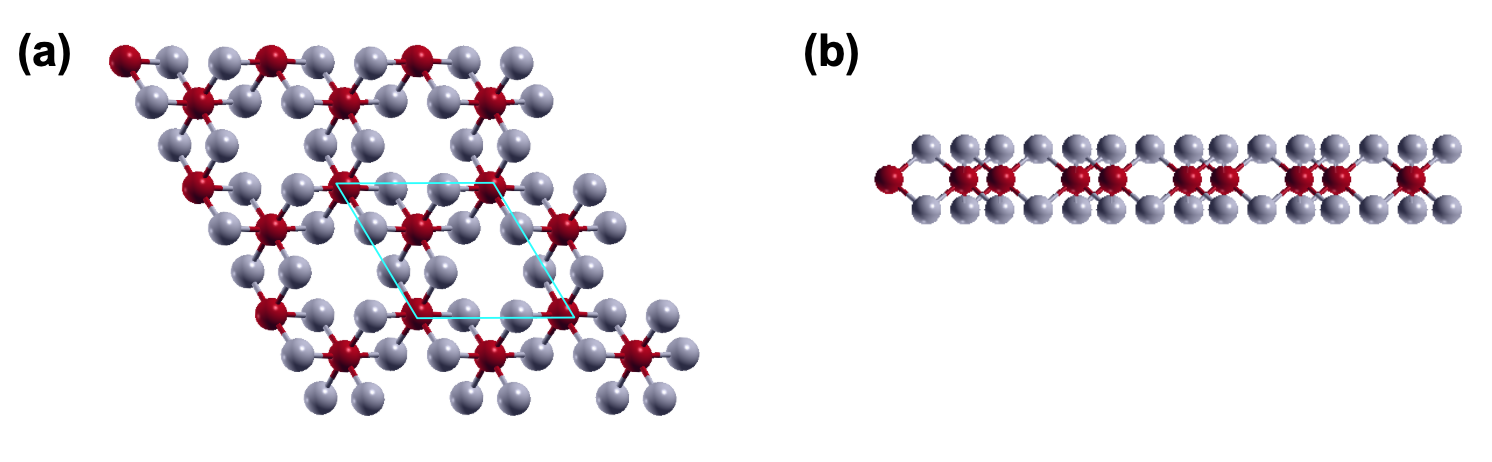}
    \caption{(a) Top view and (b) side view of optimized geometric structure of pristine \ch{VI_3}. Red and grey balls represent V and I atoms, respectively. The light blue box indicates the primitive cell of 2 V and 6 I atoms.
}
    \label{fig:figure2}
\end{figure}

\begin{figure}[htp]
    \hbox{\hspace{4.5em}\includegraphics[scale=0.4]{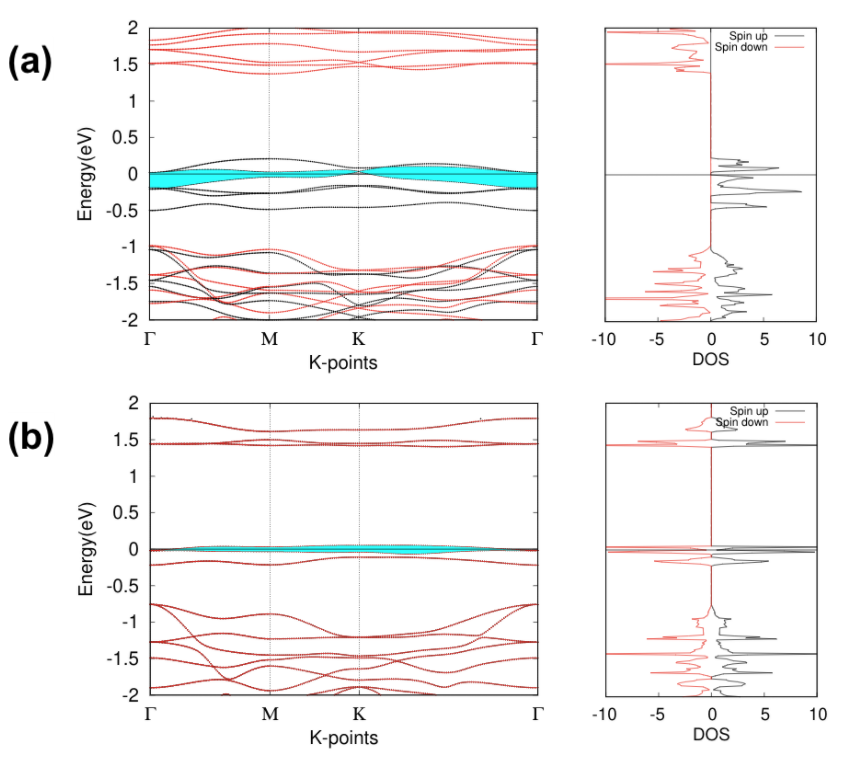}}
    \caption{Band structure and total DOS for \ch{VI_3} (a) FM state and (b) AFM state. Black represents spin up and red represents spin down. The Fermi level was set to 0 eV.}
    \label{fig:figure3}
\end{figure}

For pristine \ch{VI_3}, we used a unit cell consisting of 2 V atoms and 6 I atoms, as shown in the outlined frame in Figure \ref{fig:figure2}(a). V and its six neighboring I (each shared by two V atoms) form a \ch{VI_6} octahedron with a V honeycomb arrangement in the plane. Total energy calculations for ferromagnetic (FM) and antiferromagnetic (AFM) states of \ch{VI_3} reveal that the FM phase is the ground state. The energy difference between the AFM and FM states is 237 meV per unit cell which is comparable to 220 meV of \ch{VCl_3} \cite{zhou2016evidencing}. The calculated lattice constant, bond length, and thickness are 7.17 {\AA}, 2.76 {\AA}, and 3.18 {\AA}, respectively (listed in Table I and Table II). These values are consistent with other \ch{VI_3} calculations \cite{yang2020enhancement, baskurt2020vanadium, he2016unusual, an2019tuning}. Figure \ref{fig:figure3} shows the band structure and density of states (DOS) for \ch{VI_3} monolayer FM and AFM states. Note in Figure \ref{fig:figure3} (a),  a metallic property for the spin-up channel is observed. This can be rationalized by its two 3d electron configuration (3\ch{d^2}). With respect to the octahedral crystal field of \ch{VI_3} monolayer, the V 3d orbitals split into two-fold degenerate \ch{e_g} states and three-fold degenerate \ch{t_{2g}} states \cite{yang2020enhancement, he2016unusual}.  The two 3d electrons of \ch{V^{3+}} partially occupy the spin-up degenerate \ch{t_{2g}} states which creates the metallic property. A Dirac point is also found in the spin-up channel at the high-symmetry \textit{K} point. Meanwhile, the spin-down 3d states are almost empty with a spin down gap at 2.36 eV (listed in Table III). The total magnetic moment for the FM state of \ch{VI_3} was calculated at 4 $\mu_B$ per unit cell (see Table III). This result indicates that the four 3d electrons from the two \ch{V^{3+}} ions per unit cell predominantly contribute to the total magnetic moment of 4 $\mu_B$  while the neighboring I atom contribution is negligible. These findings are in good agreement with previous studies \cite{yang2020enhancement, he2016unusual, baskurt2020vanadium}. The imbalance between the spin-up and spin-down components is evidence of the intrinsic magnetic nature of the \ch{VI_3} material. According to Figure \ref{fig:figure1}, pristine \ch{VI_3} monolayer can be categorized as a half-metallic semiconductor (HMS), also known as a spin gapless semiconductor (SGS). 

This type of spintronic material possesses many advantages in semiconductor applications. For example, an electron from the gapless spin channel needs no threshold energy to be excited from the VBM to CBM, and the excited carrier can be fully polarized. Moreover, the carrier mobility is about two to four orders of magnitude higher than that of traditional semiconductors due to linear energy band dispersion \cite{li2016first, he2016unusual}.

\newpage

\subsection{3d TM-doped \ch{VI_3} calculations}

\subsubsection{Atomic Structure and Formation Energy}
\begin{figure}[htp]
    \centering
    \includegraphics[width=16cm]{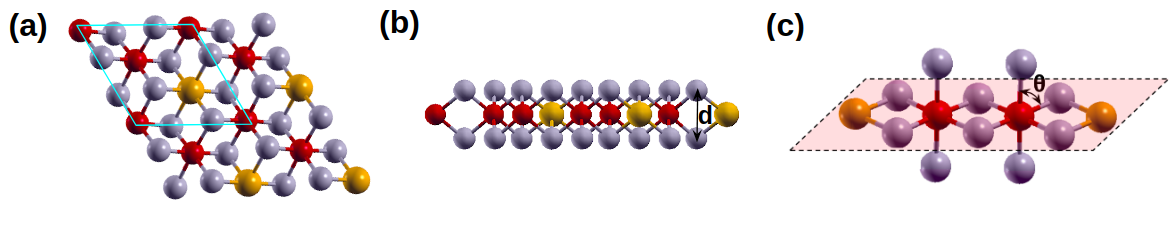}
    \caption{Atomic structure of 3d TM-doped \ch{VI_3} (a) top view, (b) side view, and (c) perspective view. Red represents V atom, grey represents I atom, and orange represents TM atom. The light blue box indicates the primitive cell of 1 TM, 2 V, and 6 I atoms. $d$ is thickness of \ch{VI_3} as defined by the vertical distance of I-I and $\theta$ is bond angle $\angle$IVI. 
}
    \label{fig:figure4}
\end{figure}

We investigated the spintronic properties of 3d TM-doped \ch{VI_3}. The selected 3d TMs were Sc, Ti, V, Cr, Mn, Fe, Co, and Ni. For TM-doped \ch{VI_3} materials, each unit cell consists of one 3d TM atom, two V atoms, and six I atoms. The TM dopant concentration was calculated at 11.1\%. As shown in Figure \ref{fig:figure4} (a), each I atom forms three bonds with one TM atom and two V atoms, whereas each I atom forms only two bonds with two V atoms for pristine \ch{VI_3} (see Figure \ref{fig:figure2} (a)). Figure \ref{fig:figure4} also shows that the transition metal atom was placed at the center of the V honeycomb. This interstitial hollow site was verified as a favorable location for dopants\cite{yang2020enhancement, baskurt2020vanadium, kong2019vi3}. In order to figure out the structural stability of 3d TM-doped \ch{VI_3} structures, we calculated the formation energy by 
\begin{equation}
    E_{form} = E_{doped} - E_{pristine} - E_{TM}
\label{eq:1}
\end{equation}
where \ch{E_{doped}} and \ch{E_{pristine}} are the total energies of the \ch{VI_3} monolayer with and without TM dopants,
respectively. \ch{E_{TM}} is the total energy of the bulk transition metal. The formation energies listed in Table II are all negative, so the formation of 3d TM-doped \ch{VI_3} is energetically favorable. 

For structural analysis (see Table I and Table II), we use pristine \ch{VI_3} as a baseline to illustrate the structure change by 3d TM doping. Both the bond length and thickness of the monolayer increased except for the bond length of Ni-doped \ch{VI_3}, which stayed the same. This may be due to the fact that the Ni atom possesses the smallest atomic radius. On the other hand, the lattice constant decreased for all dopants except for that of Ti-doped \ch{VI_3}, which increased slightly. The bond angle maintained close to 90$^{\circ}$ for all 3d TM-doped \ch{VI_3}. 

\newpage
\begin{table}[htp]
\label{tab:table1} 
\caption{Calculated bond length V-I ($b$), thickness of \ch{VI_3} as defined by the vertical distance of I-I ($d$), and bond angle $\angle$IVI ($\theta$) for pristine \ch{VI_3} and each 3d TM-doped \ch{VI_3}. See Figure 4 for the location of the bond length, thickness, and angle.}
\begin{tabular}{l c c c}

\hline
Material & $b$ ({\AA}) & $d$ ({\AA}) & $\theta$ ($^{\circ}$)\\
\hline
Pristine \ch{VI_3} & 2.76 & 3.18 & 90.1\\
\hline
Sc-doped  \ch{VI_3}& 2.84 & 3.45 & 89.1\\
\hline
Ti-doped \ch{VI_3} & 2.86 & 3.39 & 92.3\\
\hline
V-doped \ch{VI_3} & 2.85 & 3.41 & 92.7\\
\hline
Cr-doped \ch{VI_3}  & 2.84 & 3.42 & 90.8\\
\hline
Mn-doped \ch{VI_3} & 2.84 & 3.41 & 90.9\\
\hline
Fe-doped \ch{VI_3} & 2.82 & 3.43 & 93.3\\
\hline
Co-doped  \ch{VI_3} & 2.82 & 3.39 & 92.4\\
\hline
Ni-doped  \ch{VI_3} & 2.76 & 3.34 & 91.8\\
\hline
\end{tabular}

\end{table}

\begin{table}[hbp]
\label{tab:table2} 
\caption{Calculated lattice constants ($a$), TM atomic radius ($r$), and formation energy ($E_{form}$)}
\begin{tabular}{l c c c}

\hline
Material & $a$ ({\AA}) & $r$ (a.u.) & $E_{form}$ (eV)\\
\hline
Pristine  \ch{VI_3} & 7.17 & n/a & n/a\\
\hline
Sc-doped \ch{VI_3} & 7.05 & 2.4 & -3.53\\
\hline
Ti-doped  \ch{VI_3} & 7.19 & 2.3 & -1.93\\
\hline
V-doped  \ch{VI_3} & 7.14 & 2.2 & -0.60\\
\hline
Cr-doped \ch{VI_3} & 7.06 & 2.1 & -0.31\\
\hline
Mn-doped  \ch{VI_3}& 7.10 & 2.1 & -0.71\\
\hline
Fe-doped  \ch{VI_3} & 6.96 & 2.1 & -0.07\\
\hline
Co-doped  \ch{VI_3} & 7.04 & 2.1 & -0.76\\
\hline
Ni-doped \ch{VI_3} & 6.85 & 1.8 & -0.25\\
\hline
\end{tabular}
\end{table}

\newpage
\subsubsection{Band Structure and Density of States}
\begin{figure}[htp]
    \centering
    \includegraphics[width=16cm]{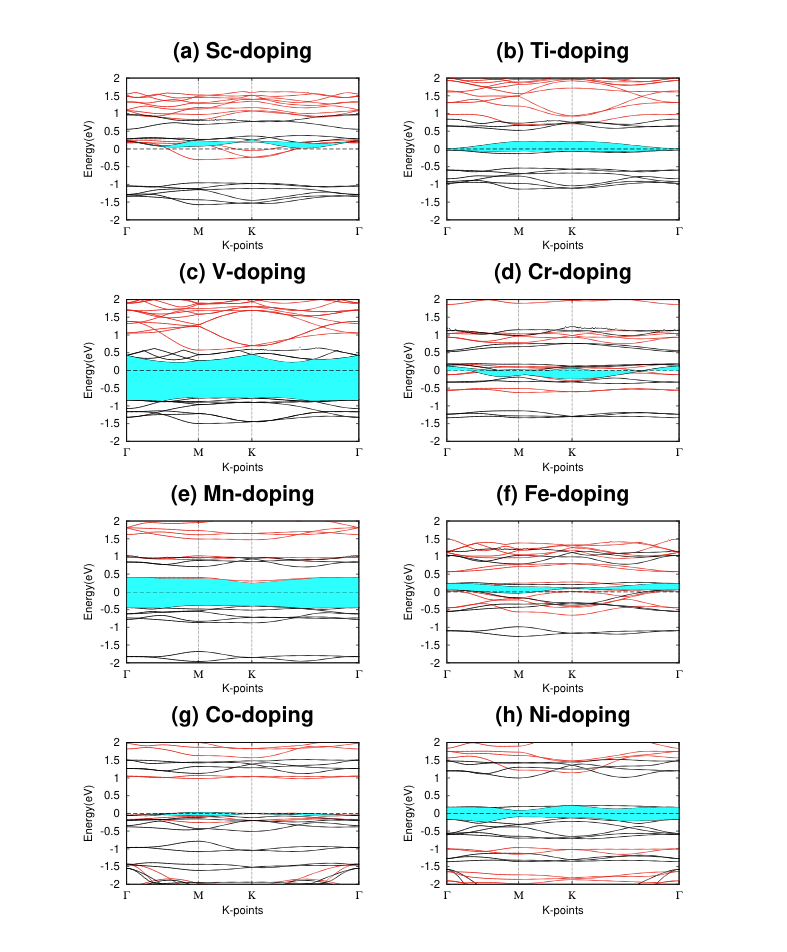}
    \caption{Band structures of TM-doped \ch{VI_3}. (a)-(h) represent the band structures for Sc-, Ti-, V-, Cr-, Mn-, Fe-, Co-, and Ni-doped \ch{VI_3}, respectively. Black bands represent spin-up bands and red bands represents spin-down bands. The Fermi level was set to 0 eV in the band structures.}
    \label{fig:figure5}
\end{figure}

Shown in Figure \ref{fig:figure5} are the band structures of all TM-doped \ch{VI_3} monolayers. Based on the band structures, the eight materials are divided into two groups: non-semiconductor spintronic materials and the desirable semiconductor spintronic materials (HMS, HSC, and BMS) according to the definition from Figure \ref{fig:figure1}.

\begin{figure}[htp]
     \includegraphics[width=12cm]{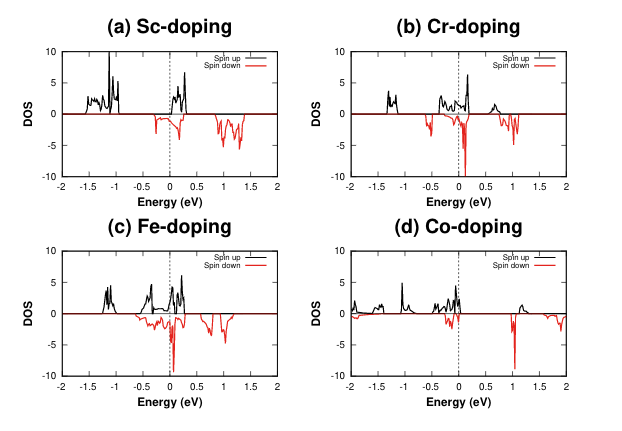}
    \caption{Total DOS of non-semiconductors (a) Sc-doped \ch{VI_3}, (b) Cr-doped \ch{VI_3}, (c) Fe-doped \ch{VI_3}, and (d) Co-doped \ch{VI_3}. Black represents the spin up states and red represents the spin down states. The Fermi level was set to 0 eV in the DOS.}
    \label{fig:figure6}
\end{figure}

Non-semiconductor spintronic materials are shown in Figure \ref{fig:figure5} (a), (d), (f), and (g). The energy bands are mixed between the spin up and spin down channels near the Fermi level for Sc-, Cr-, Fe-, and Co-doped \ch{VI_3} materials. There are no band gaps in either spin channels. In addition, their density of states (DOS) (see Figure \ref{fig:figure6}) clearly show mixed spin up and spin down channels near the Fermi Energy, indicating that Sc-, Cr-, Fe-, and Co-doped \ch{VI_3} materials are conductive and not fully spin polarized. Their total magnetic moments are calculated at 5.01 $\mu_B$, 3.40 $\mu_B$, 4.09 $\mu_B$, and 8.01 $\mu_B$ for Sc-, Cr-, Fe-, and Co-doped \ch{VI_3}, respectively (see Table III). Therefore, they are not HMS, HSC, and BMS candidates but typical ferromagnetic metals of low spin polarization with some magnetization.

Semiconductor spintronic materials with HMS, HSC and BMS properties are shown in Figure \ref{fig:figure5} (b), (c), (e), and (h). The energy bands near the Fermi Energy for Ti-, V-, Mn-, and Ni-doped \ch{VI_3} materials  are only from one spin channel, indicating full spin polarization. Specifically (see Table III), the spin up gap and spin down gap for Ti-doped \ch{VI_3} are 0 eV and 3.08 eV, respectively. Thus, Ti-doped \ch{VI_3} results in half-metallic semiconductor properties (HMS) with a total band gap of 0 eV. On the other hand, V- and Ni-doped \ch{VI_3} have spin up gaps of 1.04 eV and 0.11 eV (a narrow gap in one channel) and spin down gaps of 3.63 eV and 2.17 eV (a wide gap in the other channel), respectively. Therefore, V- and Ni-doped \ch{VI_3} possess half-semiconductor properties (HSC) with a total band gap of 1.04 eV and 0.11 eV, respectively. As for the Mn-doped \ch{VI_3} case, the spin up gap and spin down gap are 1.09 eV and 2.36 eV, respectively. Furthermore, the VBM and CBM are fully spin polarized in the opposite spin direction. Therefore, Mn-doped \ch{VI_3} exhibits an unusual bipolar magnetic semiconductor property (BMS) with a total indirect band gap of 0.62 eV. 

\begin{figure}[htp]
    \centering

     \includegraphics[width=16cm]{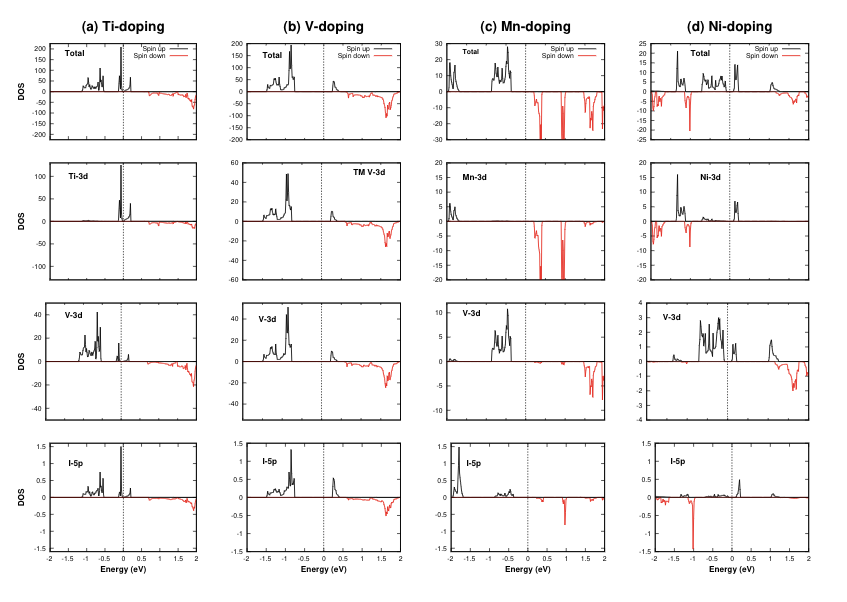}
    \caption{DOS and PDOS of semiconductors (a) Ti-doped \ch{VI_3}, (b) V-doped \ch{VI_3}, (c) Mn-doped \ch{VI_3}, and (d) Ni-doped \ch{VI_3}. Black represents the spin up states and red represents the spin down states. The Fermi level was set to 0 eV in the DOS.}
    \label{fig:figure7}
\end{figure}

To understand the origins of the HMS, HSC, and BMS characteristics, the calculated total and partial density of states (DOS) are presented in Figure \ref{fig:figure7} (a)-(d) for Ti-, V-, Mn-, and Ni-doped \ch{VI_3} monolayers, respectively. First, we analyze the Ti-doped \ch{VI_3} DOS with its HMS property (see Figure \ref{fig:figure7} (a)). Note that the DOS are plotted with different scales so that the details of each PDOS can be revealed. The total DOS around the Fermi level is mainly from the Ti 3d electrons and the V 3d electrons. There is only a small weight of the I states that contributes to the states near the Fermi level. Comparing the V 3d PDOS to the DOS of pristine \ch{VI_3} (see Figure \ref{fig:figure3} (a)), we observed that the V 3d spin up states shifted to a lower energy and lost its metallic property. The three-fold degenerate \ch{t_{2g}} states are no longer partially filled with two electrons, but instead have a full occupancy with three \ch{t_{2g}} electrons due to the lower and more stable states. This is explained by its configuration change from 3\ch{d^2} of \ch{V^{3+}} ion to 3\ch{d^3} of \ch{V^{2+}} ion with a fully occupied \ch{t_{2g}} state. In fact, a similar electronic property was observed in its sister compound \ch{CrI_3} which has a 3\ch{d^3} configuration from its \ch{Cr^{3+}} ion\cite{guo2018half, tomar2018thermoelectric, ge2019interface}. Also, after Ti doping, the unit cell of pristine \ch{VI_3} (equivalent to \ch{V_2I_6}) becomes \ch{V_2TiI_6}. Thus, the oxidation state for the V ion changes from \ch{V^{3+}} to \ch{V^{2+}} and Ti becomes \ch{Ti^{2+}} (3\ch{d^2}). Therefore, the Ti 3d spin up PDOS exhibits a metallic property because its two 3\ch{d} electron configuration (3\ch{d^2}) partially fills the spin-up \ch{t_{2g}} states. Finally, we examine the I 5p PDOS. It aligns perfectly with the V 3d electron states and Ti 3d electron states, suggesting I 5p electron orbital hybridization with the Ti and V 3d electrons to form Ti-I and V-I bonds. Overall, the Ti-doped \ch{VI_3} material shows HMS property because the spin-up 3d electron states of Ti dominate near the Fermi Energy while the spin down DOS is almost empty, indicating full spin polarization. 

For the case of V or Ni-doped \ch{VI_3} materials with HSC property as shown in Figure \ref{fig:figure7} (b) and (d), V 3d spin up states also shifted to lower states and lost their metallic property, resulting in a semiconducting behavior. We believe that the same mechanism occurs for the V atom in that its 3d electron states become fully occupied \ch{t_{2g}} states as the TM atom bonds with the I atoms. For the I 5p PDOS, we also observed a perfect alignment of the I 5p electron states with the V 3d and TM 3d electron states, indicating a strong orbital hybridization of I atoms with the TM and V atoms. 

Shown in Figure \ref{fig:figure7} (c) is the Mn-doped \ch{VI_3} with its BMS property. Note that for V 3d PDOS, the spin up states of V 3d electrons also shifted to lower states and lost its metallic property due to the same aforementioned configuration change of V atoms. In addition, the Mn PDOS confirms the opposite direction of the CBM to the VBM from the V 3d PDOS. This results in an unusual BMS property, which has been discovered in Mn related compounds \cite{du2016weak, li2014half, bannikov2013ab, li2016first}. Again, the I 5p PDOS reveals a good match with the V and Mn 3d electron states, indicating strong orbital hybridization of I atoms with Mn and V atoms to form Mn-I and V-I bonds. Notably, the strong spin down states of Mn PDOS above the Fermi Energy level have become the determining factor for the BMS property in the Mn-doped \ch{VI_3} material. 

The calculated total magnetic moments are 8 $\mu_B$, 9 $\mu_B$, 11 $\mu_B$, and 4 $\mu_B$ for Ti-, V-, Mn-, and Ni-doped \ch{VI_3} materials, respectively (see Table III). The integer values of the calculated total magnetic moments also prove full spin polarization characteristics \cite{wu2017half}. The extra 3d electrons from the dopants enhanced the magnetic moment except for that of Ni-doped \ch{VI_3} which stayed the same as the \ch{VI_3} magnetic moment at 4 $\mu_B$. The Ni configuration with eight 3d electrons may contribute to this exception. A previous study has indicated that an equal distribution of Ni 3d electrons in both spin channels results in no magnetic improvement\cite{wu2017half}. Ni 3d PDOS, as shown in Figure \ref{fig:figure7} (d), also reveals a similar contribution to both spin channels for the occupied states. On the contrary, the spin down channels are almost empty for the occupied states of Ti-, V-, and Mn-doped \ch{VI_3} materials, resulting in their enhanced magnetic moments.

Overall, the band structure and DOS calculations clearly identify that Ti-, V-, Mn-, and Ni-doped \ch{VI_3} materials are effective nanospintronic materials for semiconductor applications with HMS, HSC or BMS properties.

\begin{table}[htp]
\label{tab:table3} 
\caption{Calculated total magnetization ($Mag$) per unit cell, total band gap ($E_{gap}$), spin-up band gap ($E_{up}$), and spin-down band gap ($E_{down}$) for each material. $Mag_{TM}$ is the magnetic moment of the TM species after doping. All band gaps are direct band gaps except the values marked with ind which represents indirect band gap.}

\begin{tabular}{l c c c c c}
\hline
Material & $Mag$ ($\mu_B$) & $E_{gap}$ (eV) & $E_{up}$ (eV) & $E_{down}$ (eV) & $Mag_{TM}$ ($\mu_B$)\\
\hline
Pristine \ch{VI_3} & 4.00 & 0.00 & 0.00 & 2.36 (ind) & n/a\\
\hline
Sc-doped \ch{VI_3} & 5.01 & 0.00 & 1.04 & 0.00 & -0.42\\
\hline
Ti-doped  \ch{VI_3} & 8.00 & 0.00 & 0.00 & 3.08 (ind) & 1.53\\
\hline
V-doped \ch{VI_3} & 9.00 & 1.04 & 1.04 & 3.63 & 2.52\\
\hline
Cr-doped \ch{VI_3} & 3.40 & 0.00 & 0.00 & 0.00 & 3.57\\
\hline
Mn-doped \ch{VI_3} & 11.00 & 0.62 (ind) & 1.09 & 2.36 (ind) & 4.28\\
\hline
Fe-doped \ch{VI_3} & 4.09 & 0.00 & 0.00 & 0.30 & 3.04\\
\hline
Co-doped \ch{VI_3} & 8.01 & 0.00 & 0.00 & 0.00 & 2.14\\
\hline
Ni-doped \ch{VI_3} & 4.00 & 0.11 & 0.11 & 2.17 & -1.06\\
\hline
\end{tabular}
\end{table}

\newpage
\section{Conclusion}
In conclusion, the electronic and magnetic properties of pristine \ch{VI_3} and 3d TM-doped \ch{VI_3} monolayer
have been studied within the framework of first-principles calculations. Our investigation indicates that 3d TM interstitial doping is an effective way to enhance \ch{VI_3} spintronic property. In particular,  Ti-, V-, Mn-, and Ni-doped \ch{VI_3} hold great promise as candidates for next generation semiconductor spintronic applications with enhanced magnetizations from 4 $\mu_B$ up to 11 $\mu_B$, diverse electronic properties (HMS, HSC and BMS), and full spin polarization. Our study was limited to interstitial doping 3d TMs which is only a fraction of the possibilities for tuning the \ch{VI_3} material. In the future, research can be extended to 4d TM interstitial doping, optimized dopant concentrations, and substitutional doping. 

\section{Declaration of Competing Interest}
The authors declare that they have no known competing financial interests or personal relationships that could have appeared to influence the work reported in this paper.

\section{Acknowledgements}
The authors would like to thank Dr. Gefei Qian for the technical support.

\newpage
\bibliography{sources}{}
\bibliographystyle{ieeetr}

\end{document}